# Quantitative analysis of the dynamics of maternal gradients of the early Drosophila embryo

Victoria Yu. Samuta, Alexander V. Spirov

Sechenov Institute of Evolutionary Physiology and Biochemistry of the Russian Academy of Sciences (IEPhB RAS), St-Petersburg, Russia

## Abstract
The predetermination, formation and maintenance of the primary morphogenetic gradient (bicoid gradient) of the early Drosophila embryo involves many interrelated processes. Here we focus on a system-biological analysis of the processes of redistribution of bicoid mRNA in an early embryo. The results of a quantitative analysis of experimental data, together with the results of their dynamic modeling, substantiate the role of active transport in the redistribution of bicoid mRNA.

## Резюме
Предопределение, становление и поддержание первичного морфогенетического градиента (градиента бикоида) раннего эмбриона дрозофилы включает множество взаимосвязанных процессов. Здесь мы сосредоточимся на системно-биологическом анализе процессов перераспределения мРНК бикоида в раннем эмбрионе. Результаты количественного анализа экспериментальных данных вместе с результатами их динамического моделирования обосновывают роль активного транспорта в перераспределениях мРНК бикоида.

## Введение

Концепция морфогенетических градиентов, участвующих в развитии эмбриона была предложена в форме «физиологических» градиентов активности еще десятилетия назад (Child, 1928). В 1960-х и начале 1970-х годов эта концепция была уточнена в терминах субстратов (морфогенов), которые диффундируют через массивы клеток, формируя пространственные градиенты концентраций (Wolpert, 1969; Crick, 1970; Meinhardt, 1972). Простые молекулярные механизмы, такие как локализованный синтез белка, диффузия и деградация могут создавать в клетках зародыша постоянное неравномерное распределение морфогенов, регулирующих формирование эмбрионального паттерна. Клетки (или их ядра) способны распознавать локальные концентрации молекул морфогена, и в ответ запускать активацию определенных генов при достижении определенного порога концентрации данных молекул вещества (Wolpert, 1969; Crick, 1970). В результате начинают формироваться паттерны экспрессии генов, контролирующих последующий морфогенез. С тех пор теория была впервые разработана, ряд морфогенетических градиентов был подробно изучен экспериментально; наиболее тщательно изученным является градиент белка Bicoid (Bcd) вдоль передне-задней (AP) оси эмбриона дрозофилы (*Drosophila*).

Градиент белка Bcd возникает, когда материнская мРНК bcd накапливается (депонируется) на переднем конце эмбриона на ранней стадии его формирования (St Johnston et al., 1989). bcd мРНК начинает транслироваться как только откладывается яйцо и градиент белка возникает в течение 3-х часов. Затем морфогенетический градиент Bcd считается примерно 20 - 30 генами-мишенями, которые проявляют свое действие экспрессируя дискретные паттерны вдоль передне-задней оси (Jaeger et al., 2004; Schroeder et al., 2004; Ochoa-Espinosa et al., 2005; Manu et al., 2009a; Manu et al., 2009b; Ochoa-Espinosa et al., 2009).

мРНК гена bcd дрозофилы депонируется на кончике головной части зиготы в виде конгломерата, образуя комплекс с несколькими белками (Weil et al., 2012). Это формирует крайне специализированную структуру (P-тельца [Weil et al., 2012]), которая поддерживает устойчивость трансляции и выработку в большом количестве белка Bcd. Однако, до сих пор не выяснен окончательный механизм работы данной структуры, поэтому широко используемая рабочая гипотеза рассматривает этот конгламерат как точечный источник (не имеющий пространственной протяженности) для синтеза белка Bcd. Это привело к созданию модели Синтез-Деградация-Диффузия (SDD) для экспоненциального градиента концентрации белка Bcd. Идея модели SDD в том, что синтез белка бикоид начинается в передней точке (части) зародыша, далее белок диффундирует от головного к хвостовому концу и затем разлагается.

Другие морфогены были обнаружены и охарактеризованы после нахождения градиента белка Bcd. Эти морфогены контролируются сложными и избыточными регуляторными сетями, которые, вероятно, способствуют устойчивости к внешним возмущениям и внутренним флуктуациям, в связи с чем, SDD модель становиться подозрительно простой. Недавние исследования первичного морфргенетического градиента наводят на мысль о большей сложности его регуляции. В частности, выяснилось, что mRNA гена bcd формирует пространственный градиент, который является высоко динамичным в пространстве и времени, что имеет значительные последствия для механизма формирования паттерна морфогенетического градиента белка Bcd.

Несмотря на то, что мРНК bcd тесно связана с передним кортикальным слоем развивающегося яйца, об этом было известно с конца 1980-х годов, после оплодотворения РНК становится более диффузно распределенной (St Johnston et al., 1989). Данные Weil и соавт. [2008] показывают, что высвобождение bcd из кортикального слоя вызывается перестройкой актинового цитоскелета вследствие активации яйцеклетки. Активация яйца также вызывает перераспределение мРНК гена bcd за пределами процессирующих телец (P телец), участков цитоплазмы, где происходит регуляции трансляции и деградация мРНК (Weil et al., 2012). Weil и др. [2012] предлагают, что активация трансляции мРНК гена bcd является следствием разборки P-телец при активации яйца.

Трансляция белка *Bicoid* часто считается постоянной в течение первых 3-х часов от начала развития зародыша (когда градиент устанавливается), причем мРНК bcd остается стабильной в течение этого времени, а лишь затем начинает деградировать (Surdej and Jacobs-Lorena, 1998). Однако, длина поли (A) хвоста РНК bcd динамична, достигает максимальной длины через ~ 1.5 часа от момента начала развития; так, что синтез белка Bcd не может быть постоянным (Salles et al., 1994). Свидетельством нелинейного синтеза белка Bcd является почти экспоненциальный рост числа ядер во время формирования и поддержания градиента Bcd: большинство молекул Bcd находятся внутри ядер во время этих стадий и количество ядер удваивается каждые 8-12 мин в течение первых 12 циклов деления.

В нескольких ранних публикациях (обзор [Noll, 1993]) говориться о том, что мРНК bcd образует градиент концентрации вдоль AP оси до стадии клеточной бластодермы [Frigerio et al., 1986]. Исследования градиента белка Bcd затмили описанные ранее результаты, но проблемы с моделью SDD (например, масштабирование при изменяющейся длине эмбриона), возродили интерес к формированию паттерна мРНК бикоида. Используя наиболее чувствительный метод флуоресцентной гибридизации in situ (FISH), Spirov и соавторы (2009) показали что: (1) градиент концентрации мРНК bcd формируется вдоль кортикального слоя эмбриона к началу стадии синцитиальной бластодермы (cc10); (2) градиент экспоненциально убывает с расстоянием от переднего конца зародыша и сохраняется неизменным в течение cc10-13; (3) mRNA bcd транспортируется в апикальную ядерную периплазму во время стадии синцитиальной бластодермы; и (4) транскрипты гена bcd быстро разрушаются в течение первой трети cc14.

В работах Little и др. [2011] использован новый метод количественного определения mRNA посредствам визуализации отдельных mRNA частиц в эмбрионах для того, чтобы исследовать зависимость формирования градиента белка Bcd от концентрации, локализации, и динамики mRNA гена bcd. Исследования показали, что > 90% всей мРНК bcd постоянно присутствует в 20 % передней части эмбриона (т.е. все кроме нескольких частиц мРНК ограничены передней 20% частью яйца). Исследователи пришли к выводу, хотя протяженность пространственного распределения мРНК вносит свой вклад в формирование протеинового градиента, но сам по себе градиент концентрации мРНК не может быть причиной формирования градиента белка.

Cheung и др. [2011] показали, что мРНК гена bcd играет определенную роль в механизмах масштабирования зародыша, крупные эмбрионы содержат больше запасенной материнской мРНК, чем маленькие эмбрионы. Они [Cheung и др., 2014] также обнаружили аномальные случаи крупных эмбрионов с атипично обширным распределением мРНК bcd. С помощью секвенирования, было обнаружено, что у этих эмбрионов имеются мутации в кодирующих областях генов *Vps36* и *stau*, отвечающих за кодирование белков, связанных с транспортировкой и перераспределением мРНК bcd во время оогенеза и раннего развития эмбрионов, и показано, что эти процессы являются частью паттернинга и масштабируемости bcd.

Молекулы мРНК гена bcd очень тяжелые (тяжелее, чем молекулы белка Bcd), на протяжении оогенеза и раннего эмбриогенеза они существуют в виде крупных рибо-нуклеиновых белковых (RNP) комплексов. Такие высокомолекулярные агрегаты не могут эффективно диффундировать и, вероятно, требуют специальных механизмов транспортировки. Модель ARTS [Spirov et al., 2009] (в отличие от SDD) предполагает, что такой транспортный механизм, как, например, активный транспорт вдоль микротрубочек (MT), играет критическую роль в формировании градиента мРНК bcd в оогенезе и раннем эмбриогенезе.

Используя модифицированные методы пермеабилизации (изменение проницаемости мембраны клетки) и фиксации, для того, чтобы быстро фиксировать и сохранять кортикальные MT структуры, Fahmy и др. [2014] действительно обнаружили активность микротрубочек в ранних циклах дробления эмбриона, и показали наличие обширной сети MT исключительно в кортексе головной половины зародыша. Кроме того, исследователи показали, что aTubulin67C и ncd (кинезин-подобный минус-концевой молекулярный мотора) являются крайне важными для транспорта мРНК bcd вдоль эмбрионального кортикального слоя и установления градиента мРНК.

Новые данные о роли мРНК bcd в формировании градиента белка Bcd стимулировали ряд модельных исследований механизма формирования градиента, то есть или градиент формируется полностью за счет динамики мРНК (например, модель ARTS: Lipshitz, 2009; Спиров и др., 2009 ;. Fahmy и др, 2014) или существенную роль играет диффузия белка Bcd и его деградация (Deng et al., 2010; Little et al., 2011; Liu and Ma, 2011).

Хотя в ряде проектов по моделированию обращают внимание на детали, и процессы, еще не проверенные экспериментально (Dilao - Muraro, 2010; Kavousanakis, Kanodia, Kim, Kevrekidis, Shvartsman, 2010; Deng J, Wang W, Lu LJ, Ma J., 2010; Little, Tkacik, Kneeland, Wieschaus, Gregor, 2011; Liu, He & Ma, 2011; Dalessi, Neves, Bergmann, 2012; Shvartsman – Baker, 2012; Liu – Niranjan, 2011; Liu - Niranjan, 2012), многие эксперименты уже указывают на то, что протяженность источника мРНК имеет важное значение, и что модель SDD слишком проста.

Kavousanakis и др. [2010] предположили, что белок Bcd, скорее всего быстро захватывается ядром (оказываются в "ловушке"), и, следовательно, латеральная диффузия, вероятно имеет ограниченный эффект. Исследователи обнаружили, что обширный делокализованный источник мРНК bcd может помочь решить описанную выше проблему путем локальной трансляции белка. Little и др. [2011] обнаружили, что численное моделирование с протяженным источником мРНК bcd (соответствие их наблюдениям, что мРНК bcd сосредоточена в передних 20% паттерна) может дать более точное предсказание о паттерне белка Bcd, чем модель SDD. Liu and Niranjan [2009, 2011, 2012] проделали большую работу изучая роль, которую играет динамика мРНК в формировании белкового паттерна.

Dalessi и др. [2012] показали, что пространственно-распределенные источники, или конечного размера или нормально распределенные (как статически так и динамически) приводят к более реалистичным градиентам по сравнению с моделью SDD.

Cheung и др. [2014] использовали программу Dalessi, чтобы показать, что обширный пространственно-распределенный источник мРНК bcd, вместе с измененной динамикой белка Bcd, могли бы объяснить аномальную масштабируемость наблюдаемую в линии больших эмбрионов.

Deng и др. [2010] показали, что кортикально локализованная мРНК bcd усиливает кортикальное обогащение белка Bcd относительно сердцевины эмбриона. Это указывает на потенциальные механизмы и биологическую роль наблюдаемого базально-апикального перераспределения мРНК bcd [Спирова и др., 2009].

Эти модельные исследования показывают диапазон возможных механизмов динамики мРНК bcd и их влияние на формирование морфогенетического градиента белка Bcd. В ходе исследований было показано, что необходимы новые количественные данные и их анализ для того, чтобы протестировать модели и более детально разобраться в динамике формирования морфогенетического градиента.

Среди других, обращают на себя внимание следующие нерешенные проблемы. Во-первых, чем больше расширен сайт продукции Bcd, тем ниже может быть коэффициент диффузии для молекул этого белка. Коротким источникам требуются быстрая диффузия молекул Bcd, чтобы сформировать морфогенетический градиент, тогда как протяженный источник делает процесс формирования градиента менее зависимым от скорости диффузии молекул Bcd. Во-вторых, требуемые изменения в продукции белка Bcd могут достигаться за счет адаптивности сайта продукции Bcd. В-третьих, крупные эмбрионы должны иметь более обширные и мощные сайты синтеза белка Bcd.

Большой проблемой для сбора таких количественных экспериментальных данных является очень динамичный, изменчивый и зашумленный паттерн экспрессии мРНК гена bcd (см. [Спиров и др., 2009; Alexandrov et al., 2018]). В этой публикации мы сосредоточимся на системно-биологическом анализе процессов перераспределения мРНК бикоида в раннем эмбрионе. Результаты количественного анализа экспериментальных данных вместе с результатами их динамического моделирования обосновывают роль активного транспорта в перераспределениях мРНК бикоида.

## Результаты и обсуждение

Паттерн экспрессии раннего эмбриона мухи в основном определяется первичным морфогенетическим градиентом. От материнского организма в зиготу поступает не сам фактор белка Bcd, а только мРНК гена bcd. Ядра активированной зиготы, все вместе, в скором времени секвестрируют белок Bcd во все возрастающих количествах. Мы не знаем ни одного детального изучения организации и локализации «сайтов продукции» белка Bcd, но разумно было бы ожидать, что ЭПР окружающий ядра является основным местом, где происходит трансляция. Тем не менее в зрелой зиготе мРНК бикоида закреплена в переднем кортикальном слое яйца, тогда как ядро расположено глубоко внутри сердцевинной цитоплазмы. Два события следуют за оплодотворением: mRNA+Stau - содержащие частицы высвобождаются (1) и начинаются синхронные ядерные деления (2). Некоторые важные полномасштабные перераспределения частиц mRNA+Stau, которые возможно связаны с областью продукции белка Bcd, были выявлены в ходе нашего количественного анализа (**Рис.1-2**; [Alexandrov et al., 2018; Shlemov et al., submitted]).

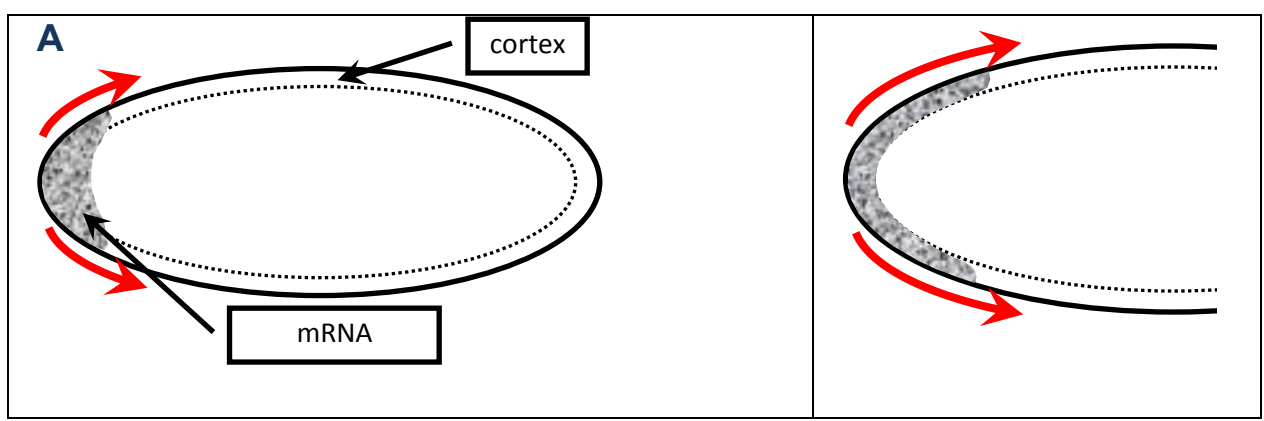

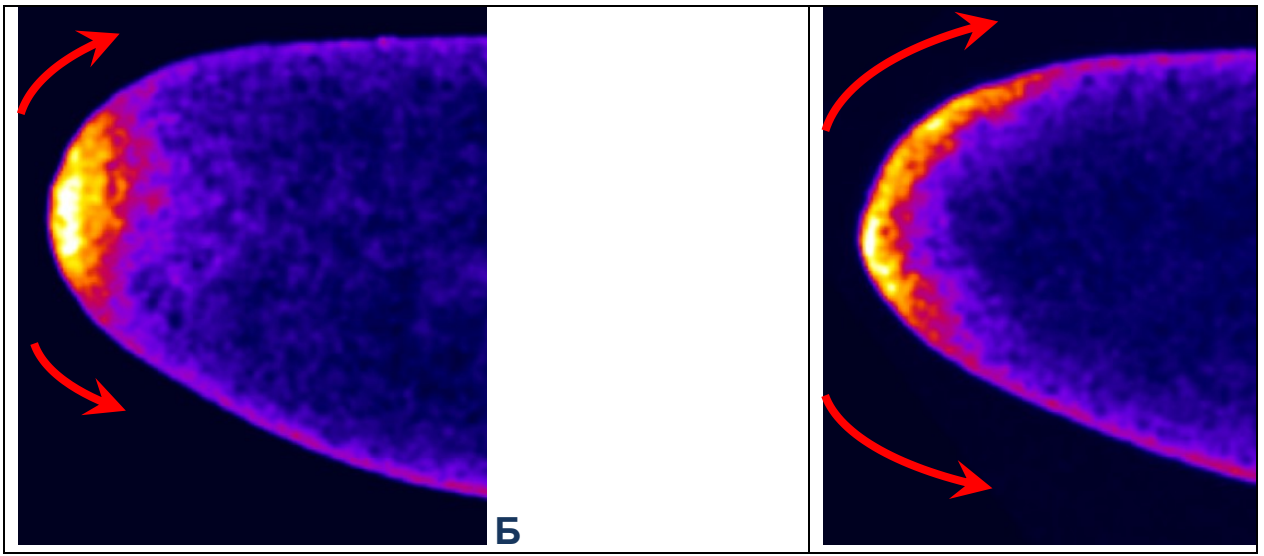

**Рисунок 1.** Два основных компонента динамики мРНК на стадии пребластодермы в передней части зародыша: движение в базально-апикальном направлении (прямые короткие стрелки красного цвета) (I) и передне-задняя миграция (изогнутые красные стрелки) (II) (мРНК обозначена как текстура). Пара рисунков (Б) иллюстрирует типичный вид скопления материала, содержащего bcd и Stau в головной части эмбриона (окраска FISH на bcd), соответствующие двум компонентам динамики схем (А).

Три ключевые особенности были выявлены в результате количественного анализа. Во-первых, это перераспределение и распространение в постериорном направлении частиц mRNA+Stau в течение первых нескольких циклов деления на стадии пре-бластодермы (до девятого цикла или 3-й стадии, примерно через 70 мин, 25 ° C) [Little et al., 2011]. Наши данные, иллюстрирующие эти события, представлены на **рис.1-2**. Деление ядер завершается их миграцией к переферии на этой стадии. Более детальный анализ профилей bcd (так же как и Stau), предполагает согласованную полномасштабную реорганизацию ранней головной области зародыша. А именно, сразу после оплодотворения частицы bcd+Stau высвобождаются и очень скоро становятся видимыми в центральной цитоплазме, формируя конус или клиновидную структуру неизвестной природы. Затем, после митоза 6, РНП (рибонуклеопротеиновый) материал возвращается обратно в передней кортикальный слой и распространяется вдоль него, формируя область, имнуемую «чашечкой» (cup). (cf [Little et al., 2011]). Ядра ассоциированные с передней обогащенной мРНК bcd областью вполне вероятно, закладывают основу для формирования высокоэффективной и регулируемой области продукции белка Bcd.

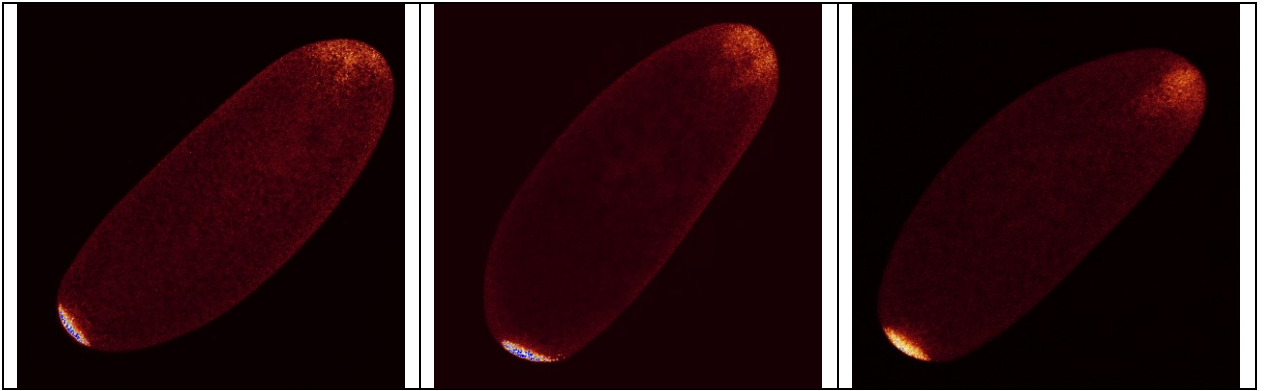

**Рисунок 2.** Переход от депо мРНК в головном кортексе яйца к источнику синтеза фактора Бикоид. События начинаются с высвобождения мРНК-содержащего материала, которое может принимать форму клиновидной или конусообразной структуры, проникающей и распространяющейся в сердцевинную цитоплазму, примерно с 4-го по 6-й цикл. Иммуноокраска на фактор Stau. На среднем изображении наблюдается пара таких структур.

Вторым событием является дальнейшее расширение области продукции белка Bcd до начала 14 деления дробления.

Третье ключевое событие – базо-апикальное перераспределение мРНК bcd (и фактора Stau) во время начала 14 цикла. Это согласуется с предыдущими наблюдениями [Спиров и др., 2009, Little и др. 2011].

Мы можем рассматривать недавно наблюдаемое перераспределение и реорганизацию мРНК-содержащего материла (bcd и Stau) как идентификацию некоторых ключевых стадий в формировании источника градиента белка Bcd (сайта продукции Bcd). Этими стадиями являются: формирование, последовательное усиление и, наконец, разборка.

Кроме того, изученные и обсужденные здесь профили интенсивности bcd и Stau и разнообразие профилей поверхностей отражают не только зашумленность данных, но и весьма динамичное поведение на каждом этапе раннего развития [Alexandrov et al., 2018; Shlemov et al., submited]. Наблюдаемые с помощью 2D-SSA тенденции отражают еще неизвестные клеточные / зиготные структуры, в то время как компоненты шума включают в себя как биологическую вариабельность и ошибки наблюдения и ошибки эксперимента. ((cf [Golyandina et al., 2012; Holloway et al., 2011; Alexandrov et al., 2008]).

### Источник формирования морфогенетического градиента

Сразу после оплодотворения частицы bcd+Stau высвобождаются и очень скоро становятся видимыми в центральной цитоплазме, формируя конус или клиновидную структуру неизвестной природы. Большая часть мРНК bcd не находится вблизи кортикального слоя, а располагается внутри эмбриона (Little at al., 2011). Затем, после митоза 6, РНП-материал возвращается обратно в передней кортикальный слой и распространяется вдоль него [Little et al., 2011]. Little с соавторами [2011] наблюдали, как ядра проникают в облако мРНК bcd во время их кортикальной миграции, к началу стадии синцитиальный бластодермы. Это обратное перераспределение мРНК bcd заканчивается распределением bcd, напоминающего чашечку, как бы покрывающую передний конец эмбриона (Рисунок S9E–H Little at al., 2011). Ядра ассоциированные с богатой bcd (мРНК) областью вполне вероятно, закладывают основу для формирования высокоэффективной и регулируемой области продукции белка Bcd.

У очень ранних эмбрионов трансляционный аппарат (который, вполне возможно, в непосредственной близости от ядер) не находится в контакте с мРНК bcd. Именно поэтому, мы сочли разумным выдвинуть гипотезу о том, что парадоксальное перемещение взад-вперед частиц РНК – способ которым мРНК bcd улавливается умножающихся в числе ядрами. И активный транспорт посредством молекулярных моторов и МТ вовлечен в данный процесс [Fahmi et al., 2014].

Следует ожидать, что сигнал Stau в головной части раннего эмбриона будет вести себя подобно bcd (так как он участвует в транспортировке комплекса bcd RNP). И изучение Stau действительно отвечает этим ожиданиям [Shlemov et al., submitted].

Наши количественные наблюдения за bcd и Stau суммируются следующим образом. Материал перераспределяется между сердцевиной головной части и ее периферией распространяется по субкортикальной цитоплазме (**Рис.1-2**), что соответствует качественным и количественным изменениям как в профилях интенсивности [Alexandrov et al., 2018; Shlemov et al., submitted].

Little с соавторами [2011] наблюдали локализацию мРНК в виде веретена или конусо-подобного распределения ,которое вдается в направлении сердцевины зародыша как видно на средне-сагиттальных планах. Авторы предполагают наличие не охарактеризованных структур, вдоль которых частицы мРНК bcd могут переноситься после активации яйца. Они заметили, что наблюдаемые частицы не продвигаться дальше в заднем направлении после n.c. 3, и наблюдения поддерживают идею, что мРНК bcd ассоциирована с подлежащими структурами цитоскелета в течение эмбриогенеза [Ferrandon et al 1994, Little at al., 2011].

Эти наблюдения за признаками некоторых клеточных структур, связанных с перераспределением мРНК, связаны с P-тельцами. Считается, что P-тельца это участки цитоплазмы, где происходит трансляция и деградация мРНК. Weil с соавторами предложил, чтобы трансляционная активность мРНК bcd является результатом разборки P телец при активации яйца [Weil и др., 2012].

Наш количественный анализ (**Рис.1-2;** [Alexandrov et al., 2018; Shlemov et al., submitted]) подтверждает предыдущие наблюдения о сложным перераспределениями bcd в пределах еще не охарактеризованной структуры [St Johnston et al., 1989; Little at al., 2011; Weil et al., 2012], которая, вероятно, основана на МТ (и микрофиламентах).

## Дальнейшее распространение области продукции белка Bcd

Как уже было отмечено ([Спиров и др., 2009] и здесь **Рис.1-2**), источник bcd распространяется постериорно далее в течении стадии синцитиальной бластодермы, начиная с последнего дробления и до 13 цикла, становясь все обширнее. Это подтверждено нашим детальным количественным анализом перераспределения сигнала от bcd и Stau в раннем эмбриогенезе [Alexandrov et al., 2018; Shlemov et al., submitted].

Мы предполагаем, что активный транспорт при помощи МТ (широко изученный в ходе оогенеза) может также работать на этих этапах (Сравни [Fahmy et al., 2014]). (К тому же, материал мог распространятся далее цитоплазматическими токами; см ниже).

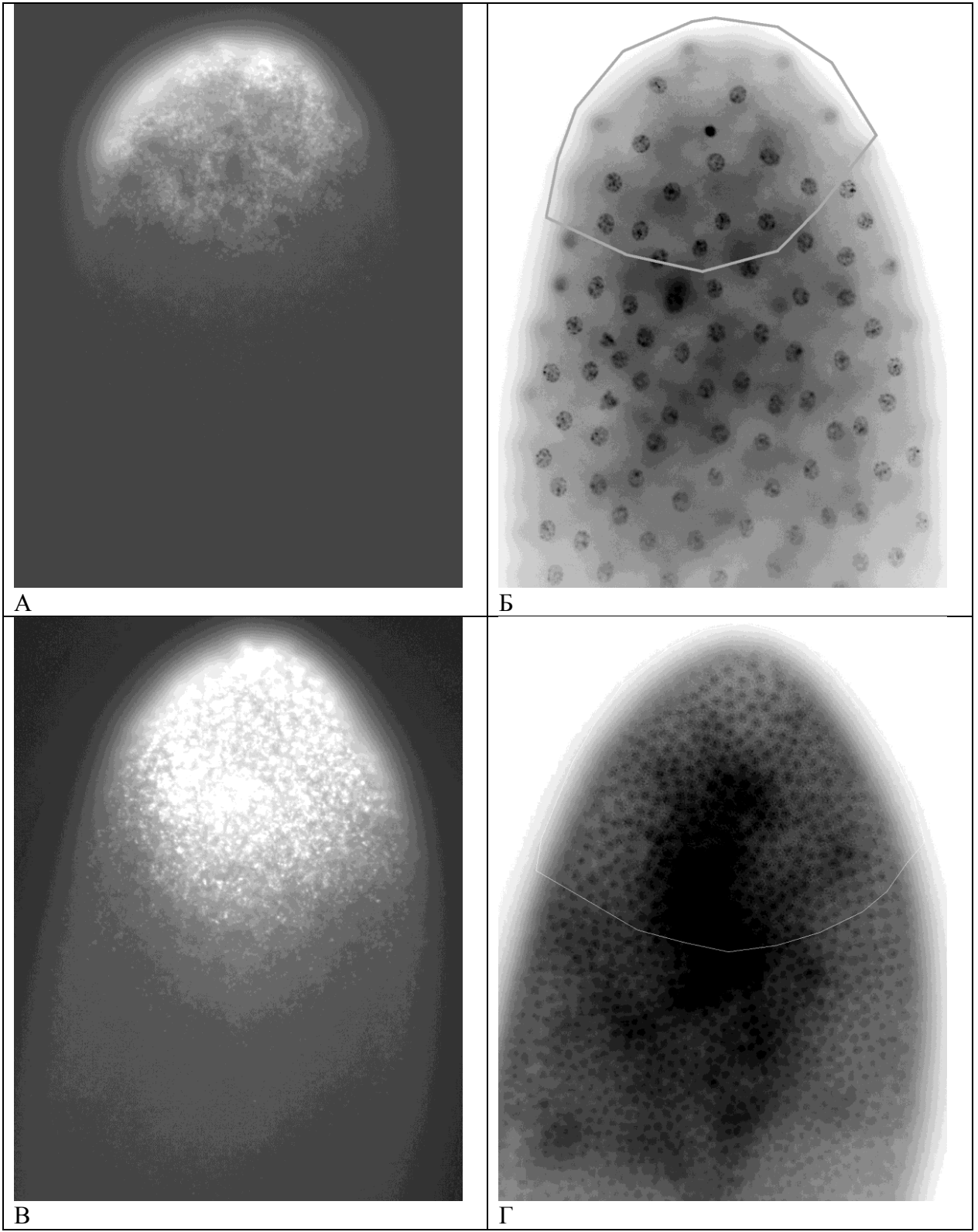

**Рисунок 3. Формирование расширенного сайта синтеза белка Bcd и дальнейшее распространение материала мРНК bcd вдоль передне-задней оси.** in situ гибридизация с дигоксигенин-мечеными РНК-зондами, FISH и эпифлуоресцентная микроскопия [Wilk et al., 2016]. (A,C) ранних и поздних (B,D) эмбрионов.

Мы интерпретируем данные, обсуждаемые нами и другими авторами, следующим образом: материал, содержащий bcd распространяется далее, чтобы включить в bcd – содержащий «cup» так много ядер, как это возможно. Мы считаем, что главной механистической причиной этого является увеличение продукции белка Bcd в сайте продукции. Эта наша гипотеза обосновывается **Рис.3**.

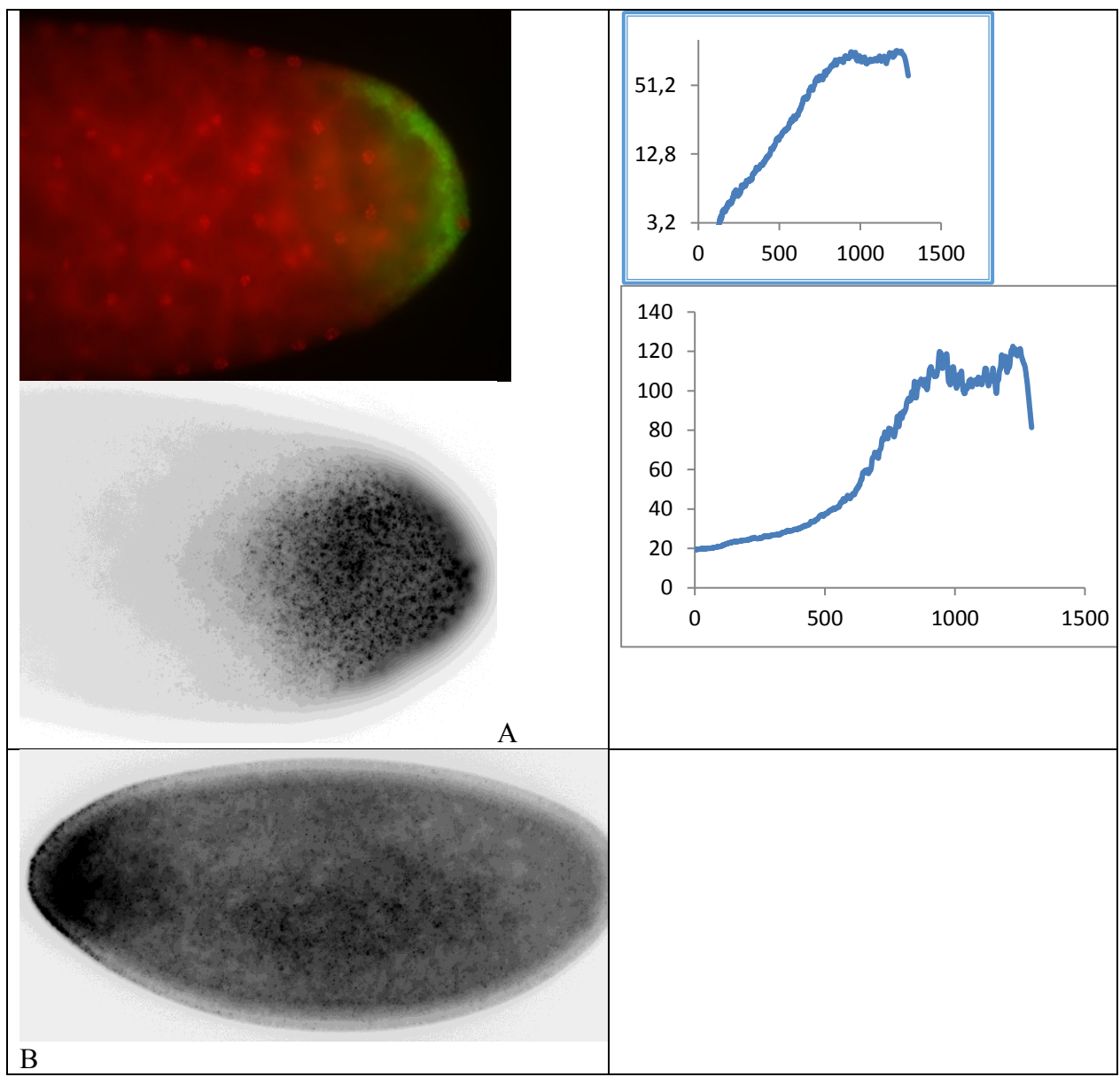

**Рисунок 4. Расширение мРНК-содержащих частиц в передне-заднем направлении в ходе 14 цикла дробления дрозофилы. in situ гибридизация с дигоксигенин-мечеными РНК-зондами, FISH и эпифлуоресцентная микроскопия (база данных FISH-fly [Wilk et al., 2016]). (А, Б) ранних и поздних (В) эмбрионов. На более зрелых эмбрионах четко видны гранулы РНК в центральной и даже хвостовой части эмбриона (данные FISH-fly).**

Результаты наблюдений, приведенные на **рис.3-4** подтверждают наш результаты и выводы **Рис.1-2**. Эпифлуоресцентная микроскопия результатов in situ гибридизации с дигоксигенин-мечеными РНК-зондами и FISH показывает как материал bcd расширяется в период от ранней синцитиальной бластодермы до 14А цикла и как увеличивается количество ядер в сайте производящем этот морфоген, как иллюстрируют **рис.3-4**.

### Динамическое моделирование раннего перераспределения мРНК бикоида

В соответствии с приведенными выше соображениями, у нас есть основания полагать, что обычно любой профиль мРНК бикоида состоит из короткого и резкого, самого антериорного градиента и длинного пологого «хвоста» [Alexandrov et al., 2018; Shlemov et al., submitted]. Хвост может быть плоским, убывающим или нарастающим (в передне-заднем направлении). Для целей динамического моделирования мы рассмотрим только профили с убывающим хвостом.

Мы предполагаем, что известные тесты с введением мРНК бикоида (RNA injection assay [[1];[2]]) отражают некоторые ключевые особенности механизмов транспорта мРНК bcd. Наша основная гипотеза здесь (требующая экспериментального подтверждения) заключается в предположении, что мРНК высвобожденная из начального антериорного конгломерата образовывает комплексы с белками-адапторами и молекулярными моторами, способные перемещаться по сети МТ синцитиального эмбриона. Следовательно, наше первое уравнение, описывающее апикальные комплексы bcd, выглядит так:

$$\frac{d[bcd]}{dt} = D_{rnp}\Delta[bcd] - C[bcd] \quad\quad\quad (Eq.\ 1)$$

Здесь [bcd]-концентрация мРНК bcd (и фактически молекулы РНК находятся в больших частично иммобилизованных макромолекулярных комплексах, которые мы для простоты называем RNP); Drnp-коэффициент диффузии для больших РНК-содержащих комплексов; C-коэффициент, соответствующий высвобождению мРНК bcd из больших медленных комплексов в более мелкие и/или существенно более подвижные (мы называем их bcd′).

Из-за очевидного очень низкой скорости распространения мРНК в постериорном направлении, коэффициент диффузии должен быть очень низким, что соответствовало бы тем крупным комплексам bcd мРНК, которые все еще, вероятно, соединены с некоторыми элементами цитоскелета.

Второе уравнение описывает динамику bcd в комплексах, способных транспортироваться по МТ:

$$\frac{d[bcd\prime]}{dt} = D_{rna}\Delta[bcd'] + C[bcd] \quad\quad\quad (Eq.\ 2)$$

---

[1] Ferrandon, D., L. Elphick, C. Nusslein-Volhard, and D. St Johnston. 1994. Staufen protein associates with the 3UTR of bicoid mRNA to form particles that move in a microtubule-dependent manner. Cell 79:1221–1232.

[2] Ferrandon, Dominique, Koch, Iris, Westhof, Eric, and Nüsslein–Volhard, Christiane, (1997) RNA–RNA interaction is required for the formation of specific bicoid mRNA 3' UTR–STAUFEN ribonucleoprotein particles, The EMBO Journal 16, 1751 - 1758.

Здесь [bcd′] - концентрация мРНК bcd в комплексах с такими моторами, как Dyenin, способными двигаться вдоль МТ; $D_{rna}$ - коэффициент диффузии для РНК-содержащих комплексов, способных к активной транспортировке по МТ.

Мы предполагаем, что МТ-сеть неориентирована, и транспортировка bcd также может быть описана законом Фика. Коэффициент диффузии в этом случае должен быть существенно выше первого (Eq. 1).

Мы выполнили достаточно простые, но достаточно подробные тесты с одномерными моделями. Мы использовали 500 ячеек по 1 мкм каждая. Время каждого теста 7200 сек, ~ 120 мин. Начальные условия были следующими: первые N клеток содержат по X единиц мРНК каждая, остальные - 0. Мы пренебрегли деградацией мРНК в течение этих первых 2 ч развития эмбриона. Модель подгонялась к реальным экспериментальным данным методом генетических алгоритмов (ГА). Типичный результат показан на **Рис.5**. Как видим, результат моделирования схож с реальными наблюдениями.

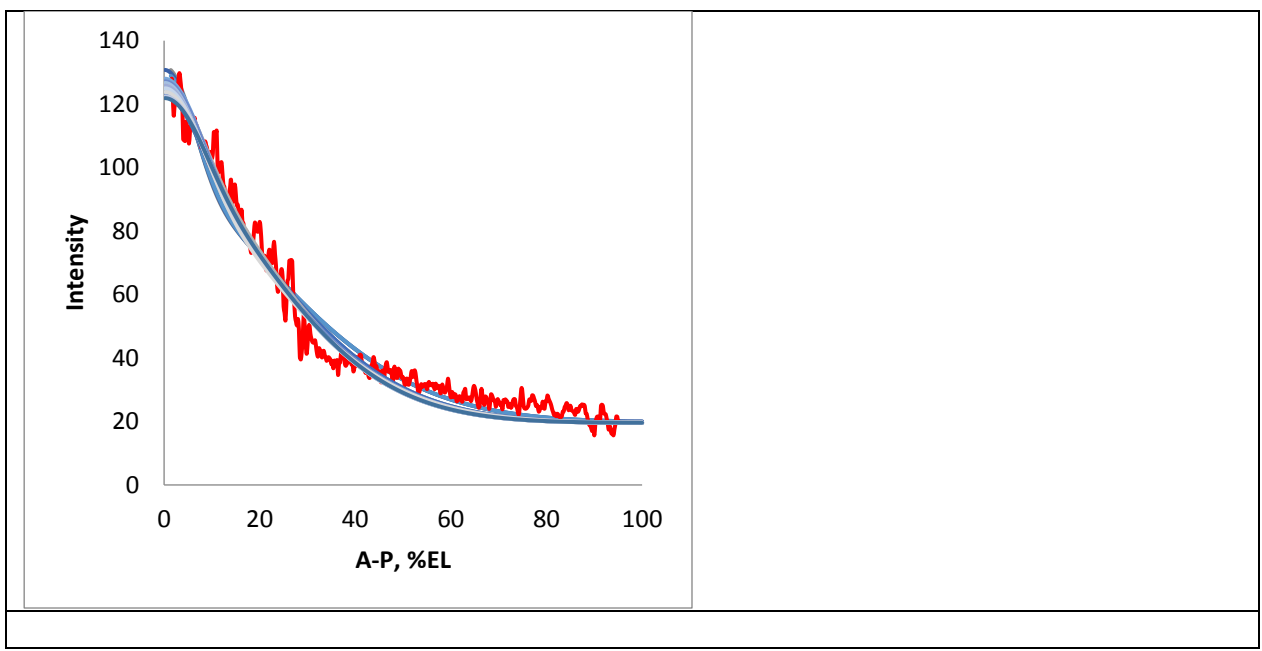

**Рисунок 5. Данные для подгонки модели и результаты моделирования к реальному репрезентативному профилю 13 цикла (суперпозиция результатов модели, голубым, и один из реальных профилей, красным).**

Как мы выяснили в этих вычислительных экспериментах, коэффициент диффузии более крупных и очень медленных частиц, содержащей bcd, составляет около 0,05 мкм$^2$ / с. Это действительно очень медленное движение, если даже мы сравним с результатами Грегора для белка Bcd, D = 0,3 мкм$^2$ / сек [Gregor et al. 2007]. Он в шесть раз меньше и соответствует либо крупным мультимолекулярным комплексам, либо частичной иммобилизации через цитоскелет, либо и тому и другому. Наоборот, как и следовало ожидать, второй процесс транспортировки выглядит действительно быстрым: D = 1,50 мкм$^2$ / сек.

Принимая во внимание, насколько велики такие комплексы, как bcd$^2$-Stau-Dynein, мы должны заключить, что этот численный результат действительно подразумевает некоторые процессы активного молекулярного транспорта для мРНК бикоида.

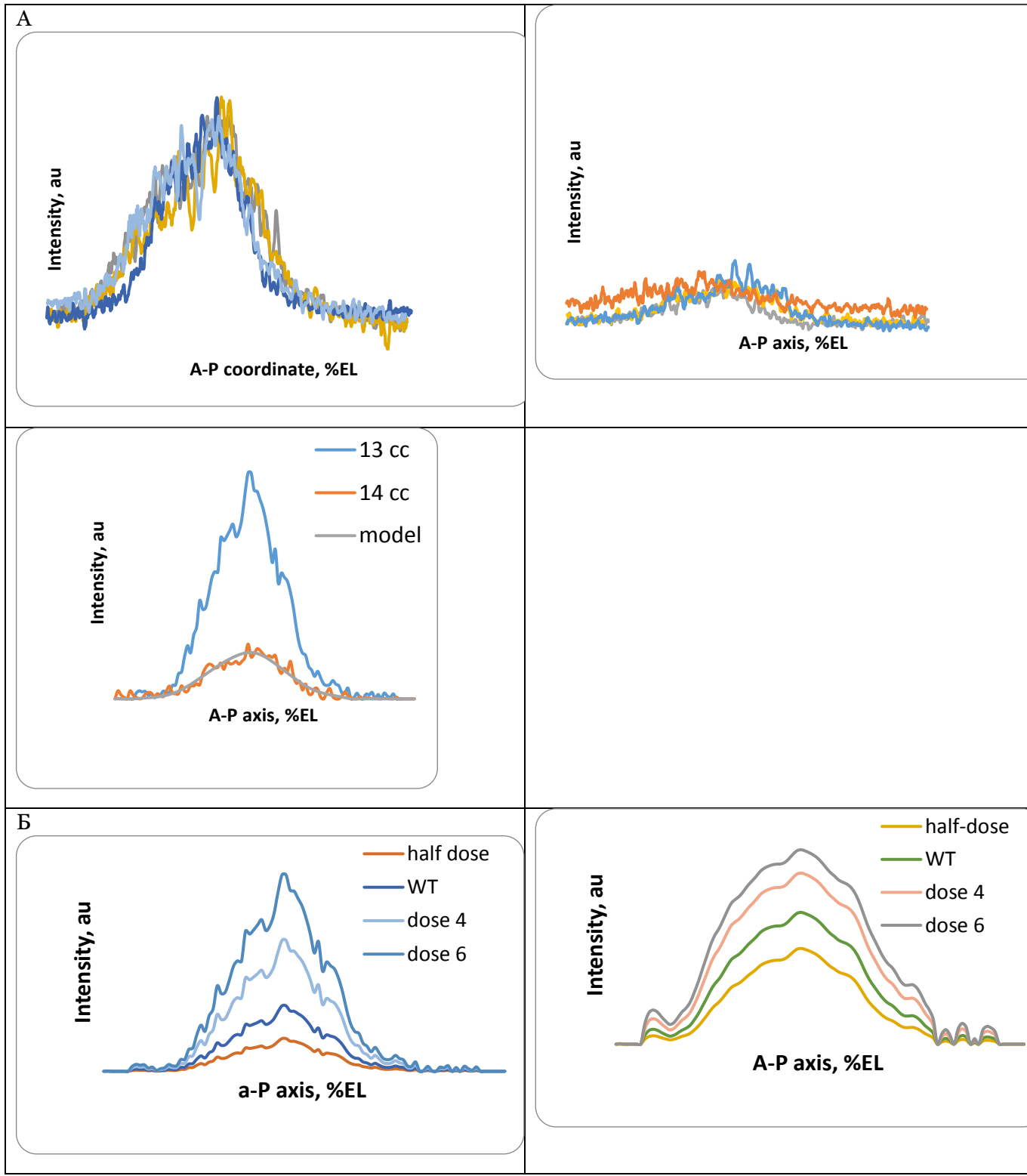

**Рисунок 6. 1D модель транспорта мРНК bcd, через 70 минут симуляции, организованная так, чтобы моделировать процесс в апикальной части раннего эмбриона (в отличие от модели предыдущего рисунка). Рандомизированный активный транспорт моделировался по закону Фика; деградацию моделировали реакцией 1-го порядка (А) или порядка 3/2 (Б). Были протестированы различные дозы**

bcd: 1 (гаплоид), 2 (WT) и более высокие дозы (4 и 6). Деградация более высокого порядка может компенсировать дозу генов на уровне мРНК.

Относительно более сложные компьютерные тесты с моделью иллюстрируются **Рис.6**. 1D модель транспорта была организованна так, чтобы моделировать процесс в апикальной части раннего эмбриона (в отличие от модели предыдущего рисунка). Здесь наша цель была подогнать результаты моделирования к реально наблюдаемым изменениям профиля bcd на стадиях зрелой синцитиальной стадии – начала целлюляризации. В качестве начальных условий здесь использовался репрезентативный профиль bcd эмбриона конца 13с цикла. Ключевым моментом здесь было изучение влияния деградации мРНК. Нас интересовали уравнение и параметры деградации мРНК (и зависимость от дозы гена). Мы осуществили поиск наилучшего значения порядка скорости деградации (первого порядка, или ниже, или выше), так чтобы получить типичный профиль эмбриона цикла самого начала 14A (~ 10 мин), начиная с заданного профиля 13 цикла. К нашему удивлению, мы обнаружили, что наилучшее значение скорости деградации мРНК bcd для соответствия экспериментальным данным было около 3/2. То есть деградация не просто пропорциональна концентрации, но чем выше концентрация мРНК, тем быстрее происходит ее деградация (**Рис.6**).

Что действительно интересно, диффузия Фикка со скоростью деградации выше единицы демонстрирует такую существенную особенность, как устойчивость решения к изменчивости начального градиента. Были протестированы различные дозы bcd: 1 (гаплоид), 2 (WT) и более высокие дозы (4 и 6). Результаты моделирования продемонстрировали, что деградация более высокого порядка (например 2/3) может компенсировать дозу генов на уровне мРНК (**Рис.6**).

Такая компенсация была отмечена на уровне точности домена фактора Hb, но механизмы все еще неясны (Houchmandzadeh, 2002). Известно, что начало цикла 14A, непосредственно перед мид-бластульным переходом, характеризуется значительно более устойчивым паттерном сегментации, чем в ранние циклы (Holloway et al., 2006; Surkova et al., 2007). То, что мы обнаружили здесь, означает, что это робастность формирования паттерна может быть реализовано не только на уровне факторов сегментации, но и на уровне некоторых мРНК.

<u>Самоусиливающийся транспорт</u>. Предположим, что крупные многомолекулярные комплексы с [bcd2-Stau-dyenin] в качестве ядра обладают не слишком высокой стабильностью у ранних эмбрионов, так что все компоненты реакций (Eq1-2) находятся в состоянии динамического равновесия. Будем называть такие комплексы зрелыми. И предположим также, что эмбрионы на этих стадиях имеют сети МТ для транспортировки комплексов в передне-заднем направлении. Это вполне обоснованная гипотеза [Спиров и др., 2009]. Если это так, квота мРНК в таких полных комплексах, готовых к транспортировке через МТ, будет сильно зависеть от суммарнй концентрации мРНК. В областях с высокими концентрациями мРНК bcd должна практически вся находиться зрелых комплексах. В то время как в областях, где количество мРНК ниже порога концентрации, она может быть в основном в неполных незрелых комплексах или даже в свободном состоянии. Следовательно, если это так, большие концентрации мРНК на переднем конце раннего эмбриона вызывают его быструю транспортировку в передне-заднем направлении.

Самоусиливающаяся деградация. Самая простая связь между количеством мРНК bcd и скоростью ее деградации, которую мы можем предположить, - это базально-апикальный активный транспорт ее на стадии клеточной бластодермы. Все соображения предыдущего раздела применимы здесь. Следовательно, можно ожидать, что чем больше количество мРНК, тем быстрее она будет транспортироваться апикально для деградации.

Самоусиливающаяся стабилизация: мРНК может повышать свою стабильность через обратные связи с повсеместно распределенными факторами, участвующими в контроле стабильности мРНК. Это может привести к длительной или даже увеличенной продукции белка, кодируемого этой мРНК.

## Что может вызывать перераспределение мРНК бикоида в самом раннем эмбриогенезе?

Детали перераспределения материала bcd mRNP, как мы это наблюдали, изображены на **Рис.7**. Это сложное движение в трех пространственных измерениях и напоминает фонтан: материал течет вперед к головному полюсу в сердцевине яйцеклетки, а затем распространяется назад (постериорно) под поверхностью головной половины яйцеклетки (**Рис.7**). Это фонтанное движение предполагает, что фонтанные силы выполняют его быстро и скоординированно.

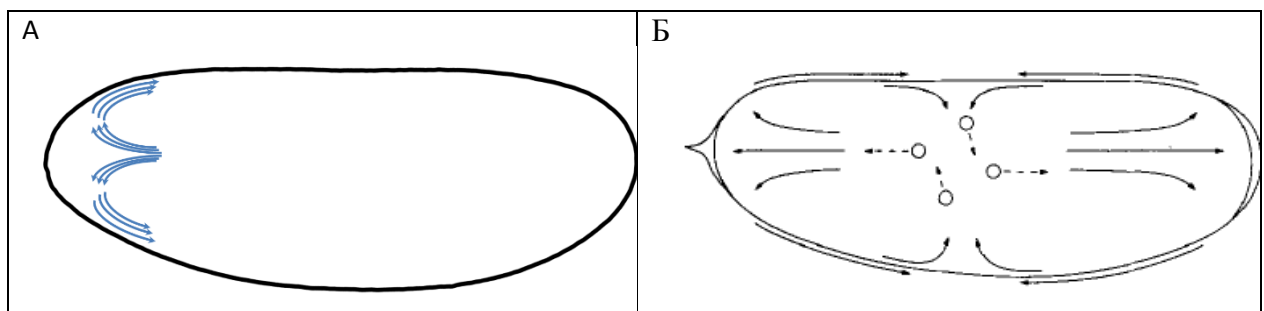

**Рисунок 7. Схема траекторий перемещения частиц мРНК бикоида в самом раннем эмбриогенезе (стадия раннего дробления, 1-6 ядерный цикл) (А) и схема фонтанных токов в эмбрионе на этой стадии (Б) [von Dassow and Schubiger, 1994].**

На самом деле, фонтанные потоки (видимо, производимые фонтанообразными силами) в головном кончике раннего эмбриона дрозофилы хорошо известны биологам. В частности, эти фонтанные токи (вместе с обсуждением сил, их вызывающих) были подробно описаны фон Дассау и Шубигером [von Dassow and Schubiger, 1994].

Итак, наша рабочая гипотеза заключается в том, что перераспределение частиц с мРНК бикоида на стадии раннего дробления связано с фонтанными токами цитоплазмы.

## Список литературы